\documentclass[10pt, conference]{IEEEtran}
\IEEEoverridecommandlockouts
\usepackage{amsmath,amssymb,amsfonts}
\usepackage{algorithmic}
\usepackage{graphicx}
\usepackage{textcomp}
\usepackage{xcolor}
\usepackage{balance}
\usepackage{booktabs}
\usepackage{caption}
\usepackage{threeparttable}
\usepackage{url}
\usepackage[nospace]{cite}

\def\BibTeX{{\rm B\kern-.05em{\sc i\kern-.025em b}\kern-.08em
    T\kern-.1667em\lower.7ex\hbox{E}\kern-.125emX}}
\begin{document}

\title{Discovering Ideologies of the \\Open Source Software Movement}


\author{
    \IEEEauthorblockN{Yang Yue\IEEEauthorrefmark{1}, Yi Wang\IEEEauthorrefmark{2}, David Redmiles\IEEEauthorrefmark{3}}
    \IEEEauthorblockA{\IEEEauthorrefmark{1}California State University San Marcos, yyue@csusm.edu}
    \IEEEauthorblockA{\IEEEauthorrefmark{2}Beijing University of Posts and Telecommunications, yiwang@bupt.edu.cn}
    \IEEEauthorblockA{\IEEEauthorrefmark{3}University of California, Irvine, redmiles@ics.uci.edu}
}



\maketitle

\begin{abstract}
Encompassing a diverse population of developers, non-technical users, and other stakeholders, open source software (OSS) development has expanded to broader social movements from the initial product development aims. Ideology, as a coherent system of ideas, offers value commitments and normative implications for any social movement, so do OSS ideologies for the open source movement. However, SE literature on OSS ideology is often fragmented or lacks empirical evidence. We thus developed a comprehensive empirical framework of OSS ideology. Following a grounded theory procedure, we collected and analyzed data from 22 OSS practitioners and 41 video recordings of Open Source Initiative (OSI) board members' public narratives. A framework of OSS ideology emerged with six key categories: membership, norms/values, goals, activities, resources, and positions/group relations; each consists of several themes. With this ideological lens, we discussed the implications and insights into the research and practice of open source.
\end{abstract}


\section{Introduction}
Open source software (OSS) is not only a software development paradigm but also a social movement \cite{Crowston12} since OSS realizes the sociological ``coming together'' of various types of individuals and organizations to form a collective identity for certain purposes on Internet-based platforms. It brings not only substantial changes in the software industry \cite{Fitzgerald06}, but also has profound implications beyond the technical realm \cite{diani1992concept}. Its scalability, continuous improvements, and community-driven innovations, offer great opportunities to champion social good and help address many challenging issues such as human trafficking, pandemic outbreaks, etc. 

Understanding a social movement should not neglect its underpinning ``heart and soul'' \cite{thompson2016society}, which is exactly the \textbf{ideology} guiding its membership and members' actions, its issue and agenda selections, its ideas about solutions to problems, and its choice of tactics. 
OSS ideology could be defined as ``\textbf{the basis of social representations regarding open source development shared by open source community}'' \cite{YueYYWR21}. Ideologies influence the OSS movement at multiple micro-, meso- and macro-levels, e.g., joining or leaving specific open source projects (micro-level), BDFL--Benevolent Dictator For Life as governance structure (meso-level), and the emergence of open innovation systems in addressing global issues (macro-level).
Thus, a comprehensive understanding of OSS ideology would inform us how the OSS movement exhibited diversities in its practices while maintaining core premises; and provide an evaluative framework for identifying the potential misfits among multiple participants' ideological orientations \cite{DanielMCH18Misfit}.  

While OSS ideologies' impacts are ubiquitous in almost every aspect of the OSS movement, a striking fact is that relatively little research focuses on it in the SE literature \cite{YueYYWR21}. Ideological elements often dispense in topics such as motivation \cite{Gerosa21motivation, sharp2009models}, and collaboration and coordination \cite{MockusFH02Coordination, JensenS05}, in a fragmented manner. One reason for the dearth of such studies might be that the term ``ideology'' is vague in its conceptualization. Researchers often took a convenient way of focusing on a narrow portion of it, e.g., the impact of ideology on effectiveness \cite{StewartG06EffectivenessIdeology}. Moreover, applying an ideological lens to investigate the OSS movement suffered from the absence of agreed-upon essential elements and an empirically-grounded theoretical framework overarching these elements \cite{YueYYWR21}.

We followed grounded theory \cite{muller2010grounded, charmaz2006constructing} to develop a comprehensive understanding of OSS ideologies. We interviewed 22 OSS participants with diverse backgrounds and compiled a dataset containing 41 Open Source Initiative \footnote{OSI is a steward organization for the OSS movement since 1998.} (OSI)'s current and former members' most recent public speeches/interviews. A substantive framework of OSS ideologies consisting of 42 themes in six broader categories emerged: \emph{Membership}, \emph{Norms/Values}, \emph{Goals}, \emph{Activities}, \emph{Position and Group-relations}, and \emph{Resources}. It extended the state-of-the-art by introducing many newly identified themes not yet covered in the SE literature. We further discussed the theoretical insights and practical implications of OSS practices around its ideologies.

\section{A Brief Review of OSS Ideologies}

Researchers have been studying the OSS movement and exploring its ideology since its inception. Ljungberg defined OSS ideology from two dimensions, i.e., zealotry, and hostility towards commercial software \cite{Ljungberg00}. Stewart and Gosain developed the three-tenet  (beliefs, values, \& norms) framework by combining the literature and famous OSS advocates' narratives \cite{StewartG06EffectivenessIdeology}. However, they admitted that their framework was still too preliminary.
Researchers have also applied these frameworks to study OSS ideology's impacts, e.g., 
the misfit of OSS ideology between contributors and projects could influence contributors' commitment \cite{DanielMCH18Misfit}. 

Another stream of research focused on particular aspects of OSS ideology, e.g., the individual motivations to participate in open source development \cite{Gerosa21motivation}, monetary incentives' effects in motivating contributors \cite{ZhangWYZLW22Sponsorship}, individuals' activities in OSS projects were also studied by researchers \cite{GeigerHI21Maintainer}, contributors' improper or unethical behaviors \cite{QiuDetecting}, etc.
Moreover, many project/community level phenomena are also related to ideology. As the fundamental element of ideology, values continuously received researchers' attention, e.g., transparency \cite{Dabbish12transparency}, diversity \cite{Bosu19diversity, Vasilescu19diversity, Canedo20women}, etc. From a structural perspective, researchers found that different roles of contributors consist of a centralized, layer-upon-layer structure \cite{JensenS07Role}, with pathways to allow leadership to emerge \cite{Trinkenreich20roles, HergueuxK22Leadership}.

In general, the research around OSS ideology were increasing \cite{Crowston12, YueYYWR21}. However, most studies tended to be knowledge fragments that focus on particular aspects, such as \emph{gender} and \emph{fairness} \cite{YueYYWR21}, and there were few efforts to integrate them into a coherent body. Moreover, they often focused on high-profile OSS projects and technical icons, and almost half of them were skewed toward a single project \cite{Crowston12}. That potentially overlooked the voice of grassroots and non-technical contributors in the OSS movement \cite{YueYYWR21}. Therefore, we collected data from both first-line OSS participants and leaders, and built an empirical framework of OSS ideologies.


\section{Methodology}

The study design followed the grounded theory methodology \cite{muller2010grounded, charmaz2006constructing} due to the lack of theoretical and empirical evidence related to OSS ideologies. An overview of the whole research process is provided in the supplemental materials\footnotemark.

\footnotetext{Supplemental materials: \url{https://doi.org/10.6084/m9.figshare.28149158}}

\subsection{Data}
\subsubsection{Data Sources}
We had two data sources: (1) interviews with OSS practitioners, and (2) video-recorded public interviews/speeches of current and former OSI board members. Both constituted legitimate data sources since grounded theory accommodates ``interviews, field observations, documents, video, etc.'' \cite{corbin2008basics}. They were not randomly chosen. We noticed that most interviewees were technical staff whose experiences were hardly beyond their projects. Thus, such a sample was insufficient to develop a holistic understanding of OSS ideologies. Meanwhile, the potential informants representing non-technical aspects of OSS (e.g., legal counsels) were often hard to access directly for interviews. Therefore, we resorted to secondary data sources. We chose the video recordings of public interviews/speeches of OSI board members on video-sharing sites, e.g., \textsc{Youtube}. They were often renowned contributors or highly-respected proponents of the OSS movement, e.g., Josh Berkus, a major contributor to Linux and PostgreSQL.

\subsubsection{Data Collection}
The data collection started with initial purposive sampling. We recruited participants and had interviews. Then we had another iteration of data collection, theoretical sampling, based on the collected data. The theoretical sampling included recruiting more participants for interviews, and collecting secondary data. We recruited 22 interviewees through multiple channels, i.e., direct emailing \textsc{GitHub} users, posting recruitment ads on social media, and using researchers' personal connections. In total, we recruited 22 participants (\textbf{P1}-\textbf{P22}) from 14 countries. They had diverse backgrounds in jobs, experiences, and other demographic factors. Video data was collected after the first wave of interviews started. Note that we decided to use OSI board members' public interviews/speeches on video-sharing sites (mostly \textsc{Youtube}) as the secondary data source. First, we compiled a list of 51 people, including both 10 current OSI board members and 41 emeritus members from OSI's website. Then, we searched online for their most recent (within five years) public interviews/speeches related to OSS. In total, 41 had such online videos. These videos were transcribed for analysis. More information is available in the supplemental materials\footnotemark[\value{footnote}].

\subsubsection{Data Analysis}
The data analysis process started immediately after some data was collected. Both interview data and secondary video data were treated equally. The results emerged during data analysis, and then formed the basis of further theoretical sampling. Therefore, we need to go back and forth between data collection and data analysis. The iterative process of data analysis and data collection stopped when \textit{Theoretical Saturation} was reached \cite{muller2010grounded}. As a typical grounded theory study, the analysis went through different phases, including \emph{Open Coding}, \emph{Axial Coding}, \emph{Selective Coding}, and \emph{Theoretical Memo Writing and Concept Refinement}. 

Moreover, we reused a pre-existing theoretical framework proposed by van Dijk \cite{VanDijk1998} to facilitate the data analysis \cite{charmaz2006constructing}. We discussed the emerging categories and concepts, sorted them, and mapped them into the framework. We paid particular attention to ensuring that the framework would be adapted to fit these categories and concepts well \cite{charmaz2006constructing}. Finally, the empirical framework of OSS ideology emerged.

\section{Results \& Findings}

\begin{figure*}[!h]
        \includegraphics[width=0.92\textwidth]{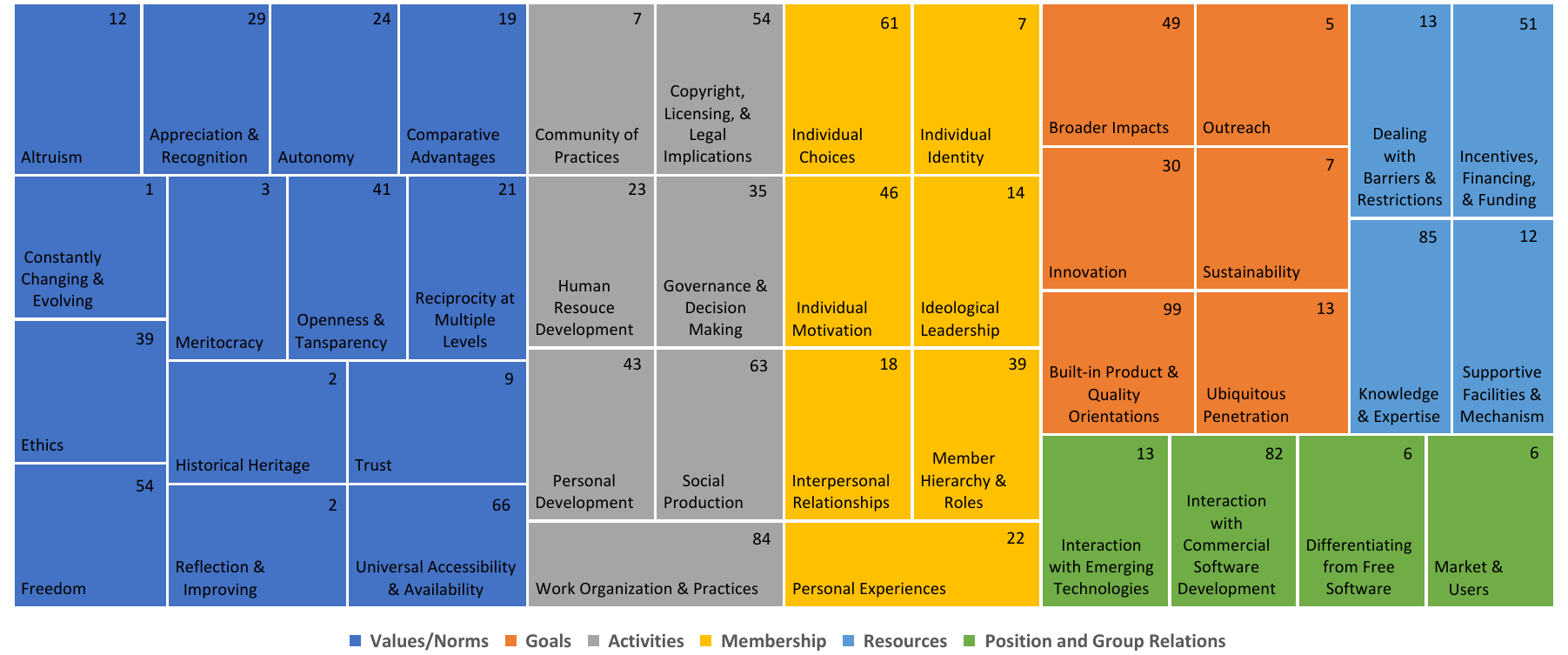}
     \centering
     \caption{The overview of the empirical framework of OSS ideology (the numbers are theme frequencies in the collected data).}
     \vspace{-1.5em}
    \label{fig:theory}
\end{figure*}

Fig. \ref{fig:theory} provided an overview of the empirical framework that emerged from the data. It fits well with van Dijk's six-category framework \cite{VanDijk1998} for ideologies, and each category contained a set of themes that emerged in the data analysis, the code frequency of each theme was also listed. In total, the theory consisted of 42 themes in six categories.

\subsection{Membership}

\textbf{Membership} defines the people involved in OSS, i.e., where they come from, why and how they join, etc. In general, ``\textit{people with the same interests}'' (\textbf{P18}) are welcome to join OSS of their free will. Seven themes are identified: (1) \emph{Individual Choices} indicates that people's memberships and actions are of their own choice, e.g., ``\textit{I enjoy spending that free time on hacking an open source project.}'' (\textbf{P12}). (2) \emph{Individual Identity} refers to people's representation of self in constructing themselves as OSS members. (3) \emph{Individual Motivation} reflects what motivates them to contribute, including intrinsic and extrinsic ones \cite{Gerosa21motivation}. (4) \emph{Ideological Leadership} is about the connection between a shared set of OSS ideologies and leadership in communities. (5) \emph{Interpersonal Relationships} describes the relationships among members, mostly in a friendly and professional manner, while conflicts also arise sometimes, since ``\textit{conflict is a natural and inevitable outcome.}'' (\textbf{Tracy Hinds}) (6) \emph{Member Hierarchy \& Roles} identifies the hierarchical structure of contributors and their privileges and responsibilities defined by their roles in the membership pyramid. (7) \emph{Personal Experiences} captures personal feelings in OSS, e.g., ``\textit{fun to collaborate}'' (\textbf{P12}) or ``\textit{a feeling of being honored}'' (\textbf{P20}).

\subsection{Values/Norms}

\textbf{Values/Norms} are defined as guiding principles in people's lives. In addition to the functions at the individual level, human values/norms' socio-cultural nature makes them be shared, known, and applied by members in a variety of OSS practices. Our data analysis revealed 14 related themes (Tab. \ref{tab:norms}).

\begin{table*}[h!]
    \centering
     \caption{The themes related to Values/Norms.}
    \begin{tabular}{p{0.15\textwidth}p{0.82\textwidth}}
    \toprule
     \textbf{Themes}    & \textbf{Meanings} \\
    \midrule
      Altruism   & OSS movement is driven by \emph{Altruism} because contributors spend their own time and effort to make something good for others voluntarily. E.g., \textbf{Baytiyeh \& Pfaffman} once wrote ``\emph{Open source software: A community of altruists}''\\
     Appreciation \& Recognition & Every contribution should be recognized. People should ``\textit{acknowledge and respect}'' (\textbf{Deborah Bryant}) altruists' contributions and give them proper credit. Moreover, ``\textit{recognition is not only about the financial reward.}'' It could be a simple ``thank you'' and ``\textit{mentioned the name on the issue of the pull requests}.''\\
     Autonomy& OSS members are self-governed and self-determined in making choices and decisions, e.g., ``\textit{there is not really anyone in open source who tells you, you should be doing this. It is very like people self-determined.}'' (\textbf{Martin Michlmayr})\\
      Comparative Advantages& OSS had \emph{Comparative Advantages} over other types of software in its members' mind, such as ``\textit{more secure by having the code fully available}.'' (\textbf{P16})\\
     Constantly Changing \& Evolving & OSS community ``\textit{changes over time}'' (\textbf{P1}), and should be evolving all the time.\\
     Ethics& Practitioners believe that ``\textit{open source is the pragmatic approach to ethics}'' (\textbf{Simon Phipps}), and OSS communities should embrace diversity and inclusion as their essential values, ``\textit{it is crucial we continue to build spaces where all people are welcome, regardless of race, gender, or sexuality}.'' (\textbf{Aeva Black})\\
     Freedom& OSS also values software freedom, as free software movement, but in a more practical way. Practitioners have full freedom, e.g., ``\textit{I can start whenever I want, I can stop whenever I want,}'' (\textbf{P12}) and ``\textit{you can do what you want.}'' (\textbf{P21})\\
     Historical Heritage& OSS should respect the historical heritages of its predecessors, i.e., the Linux Foundation, and the free software movement. (\textbf{P5})\\ 
     Meritocracy& Power should be assigned according to merit only, which was determined by ``\textit{the value of your contribution} (\textbf{Jim Jagielski}).''\\
     Openness \& Transpare-ncy & The values \emph{Openness \& Transparency} are endogenous to OSS, and should be honored in every aspect of OSS. 
     \\ 
     Reciprocity at Multiple Levels& Reciprocal expectations are prevalent, and bring mutual benefit for both individuals and projects in OSS community, e.g., ``\textit{how other people help me, I want to help other people [in the same way]}.'' (\textbf{P14}) 
     \\
     Reflection \& Improving & OSS projects should be able to self-reflect and improve their practices continuously, ``\textit{there is potential for self-correcting.}'' (\textbf{P5})\\
    Trust & OSS members should maintain certain levels of mutual trust towards each other to ensure cooperation. In OSS community, ``\textit{When someone has built up trust and has full trust... It is found efficient and has been working really well}'' (\textbf{P5}), for example, ``\textit{they know if you do something, it is going to be high-quality.}'' (\textbf{Martin Michlmayr})\\
     Universal Accessibility \& Availability & First, practitioners believe that ``\textit{source code access was a utilitarian good,}'' (\textbf{Luis Villa}) i.e., the source code should be publicly available, and anyone can access without asking for permission. Then, no one needs permission to use, modify, or distribute it.\\
     
     \bottomrule
    \end{tabular}
   \vspace{-2em}
    \label{tab:norms}
\end{table*}

\subsection{Goals}

\textbf{Goals} described what members want to achieve or realize in the OSS movement. We identified six themes. (1) \emph{Broader Impacts} captures the goals beyond software productions, such as ``\emph{brings humanity forward}'' (\textbf{P9}) in our society. (2) \emph{Built-in Product \& Quality Orientations} is the fundamental goal. Without high-quality products, no other goals could be realized. It is considered to be a key component in Stewart \& Gosain's OSS ideology model \cite{StewartG06EffectivenessIdeology}. (3) \emph{Outreach} describes the goal of promoting itself to reach broader populations, e.g., ``\textit{I tried to make it [an OSS project] heat and known by community.}'' (\textbf{P20}) (4) \emph{Sustainability} refers to OSS communities' goal of achieving sustainable dynamics and growth. (5) \emph{Innovation} could be driven by OSS development, moreover, ``\textit{it lets you innovate without having to ask anyone's permission.}'' (\textbf{Simon Phipps}). (6) \emph{Ubiquitous Penetration} reflects the OSS movement's goal of penetrating every aspect of modern society. Many practitioners expect that OSS would ``\emph{be foundational to modern technology}'' and thus become ``\emph{a way of life}.''

\subsection{Activities}

\textbf{Activities} deal with questions such as ``what do OSS contributors do?'' and ``what are expected activities in OSS?'' It is the most important one among the six categories. According to van Dijk (\cite{VanDijk1998}, pp. 70-71), an ideology system could be identified by one particular category. For open source ideology, its distinctions mainly lie in the activities, particularly the copyright \& licensing activities that define it\footnote{\emph{The Open Source Definition}, v1.9, available at \url{https://opensource.org/osd}}. Thus, OSS ideology is typically an (\textbf{activity}) ideology representing that OSS contributors loosely gather to form communities for producing software under specific OSS licenses. There are seven themes identified. (1) \emph{Community of Practices} refers to OSS community members' collective activities as a group of people who ``share a concern or a passion for something they do and learn how to do it better as they interact regularly''\cite{Krishnaveni12}. (2) \emph{Copyright, Licensing, \& Legal Implications} refers to the legal practices in OSS development, particularly about dealing with copyright and patent, using OSS license, etc. These legal activities provide explicit guarantees on the aforementioned values/norms. (3) \emph{Governance \& Decision-making} refers to governance structures, i.e., benevolent dictator, walled garden, and true meritocracy, as well as the decision-making processes corresponding to them. (4) \emph{Human Resource Development} is about developing a workforce, including attracting, recruiting, and retaining contributors, and facilitating their growth. (5) \emph{Personal Development} is the activities related to an individual's professional development, such as enriching portfolio, earning skills, and seeking career opportunities. (6) \emph{Social Production} is about the collaborative nature of its members' activities which features the collective efforts of multiple entities. (7) \emph{Work Organization \& Practices} are the practices about how work is organized in OSS development, e.g., communication and coordination, and development routines.

\subsection{Resources}

\textbf{Resources} are essential for a community to survive and develop. We identified four themes: (1) \emph{Dealing with Barriers \& Restrictions} mentions the resources that individuals and a community used to deal with challenges they faced. A typical individual-level resource is certain personality traits, e.g., resilience, for dealing with frustrations, particularly in one's early career phase \cite{SteinmacherGCR19}. (2) \emph{Incentives, Financing \& Funding} is about money. While contributions are voluntary, healthy cash flow is still important for many projects, especially large ones, to maintain community infrastructures, e.g., paying for project communication services. (3) \emph{Knowledge \& Expertise} is about the knowledge and expertise needed. Moreover, they are not only cognitive or epistemological but also involve in many social dimensions. Power is based on knowledge and makes use of it; on the other hand, power reproduces knowledge by shaping it in accordance with its anonymous intentions. 
(4) \emph{Supportive Facilities \& Mechanisms} refers to the supportive infrastructures implemented by open source communities.

\subsection{Position and Group Relations}

\textbf{Position and Group Relations} deal with a series of questions such as, \emph{what is our social position? who are our opponents? who are like us, and who are different?} For this category, there are four key themes. (1) \emph{Interaction with Emerging Technologies}  reflects the fundamental positions of OSS in enabling emerging technologies (e.g., AI and Cloud) and forming Internet-wide infrastructures together with these technologies. (2) \emph{Interaction with Commercial Software Development} and (3) \emph{Differentiating from Free Software} are about OSS' relationships with commercial software and free software. (4) \emph{Market \& Users} summarizes the relationships among OSS, market, and users from multiple perspectives such as economics, management, HCI, etc. 
\section{Threats to Validity}

Regarding the \textit{theoretical validity}, we are confident that there is no severe threat to it, partially due to the application of van Dijk’s theoretical framework. We reused its core concepts, which are well-established in a huge body of literature. The data grounded the extensions to the concepts well. We developed theorized relationships among the concepts in the context of open source development. These relationships maintain high coherence in understanding and explaining the phenomena reflected in our participants’ narratives, thus fitting the data. Regarding the \textit{external validity}, we compiled a diverse sample, including both grassroots OSS contributors and OSI board members, to reflect the population of OSS. But it was still impossible to claim that the sample could represent the whole OSS community. Thus, our framework may not be directly generalizable to all OSS practitioners and projects.

\section{Future Plans}

Our future plans include further investigations of OSS ideologies and the development of an assessment framework. First, we plan to conduct investigations of the newly identified themes in our framework. With insights into those themes, we could further develop a comprehensive understanding of OSS ideologies. Then, our framework could also serve as \emph{an assessment framework} for OSS practitioners to investigate if the desirable ideologies were upheld and to what extent. Researchers have revealed that the misfit between contributors and projects could influence OSS practices \cite{DanielMCH18Misfit}. A practical assessment system could help identify the misfit, avoid potential conflicts in OSS development, and help match projects and contributors with similar OSS ideologies, thus having high necessity. Our framework set a solid foundation for developing an assessment system for ideologies. Moreover, ideologies are ever-changing, we will also take a dynamic perspective to follow their evolution over time.


\balance
\bibliographystyle{IEEEtran}
\bibliography{references}

\begin{thebibliography}{10}
\providecommand{\url}[1]{#1}
\csname url@samestyle\endcsname
\providecommand{\newblock}{\relax}
\providecommand{\bibinfo}[2]{#2}
\providecommand{\BIBentrySTDinterwordspacing}{\spaceskip=0pt\relax}
\providecommand{\BIBentryALTinterwordstretchfactor}{4}
\providecommand{\BIBentryALTinterwordspacing}{\spaceskip=\fontdimen2\font plus
\BIBentryALTinterwordstretchfactor\fontdimen3\font minus \fontdimen4\font\relax}
\providecommand{\BIBforeignlanguage}[2]{{%
\expandafter\ifx\csname l@#1\endcsname\relax
\typeout{** WARNING: IEEEtran.bst: No hyphenation pattern has been}%
\typeout{** loaded for the language `#1'. Using the pattern for}%
\typeout{** the default language instead.}%
\else
\language=\csname l@#1\endcsname
\fi
#2}}
\providecommand{\BIBdecl}{\relax}
\BIBdecl

\bibitem{Crowston12}
K.~Crowston, K.~Wei, J.~Howison, and A.~Wiggins, ``Free/libre open-source software development: What we know and what we do not know,'' \emph{{ACM} Comput. Surv.}, vol.~44, no.~2, pp. 7:1--7:35, 2012.

\bibitem{Fitzgerald06}
B.~Fitzgerald, ``The transformation of open source software,'' \emph{MIS Quarterly}, vol.~30, no.~3, pp. 587--598, 2006.

\bibitem{diani1992concept}
M.~Diani, ``The concept of social movement,'' \emph{The Sociological Review}, vol.~40, no.~1, pp. 1--25, 1992.

\bibitem{thompson2016society}
W.~E. Thompson, J.~V. Hickey, and M.~L. Thompson, \emph{Society in Focus: An Introduction to Sociology}.\hskip 1em plus 0.5em minus 0.4em\relax Rowman \& Littlefield, 2016.

\bibitem{YueYYWR21}
Y.~Yue, X.~Yu, X.~You, Y.~Wang, and D.~F. Redmiles, ``Ideology in open source development,'' in \emph{Proc. CHASE '21}.\hskip 1em plus 0.5em minus 0.4em\relax {IEEE}, 2021, pp. 71--80.

\bibitem{DanielMCH18Misfit}
S.~L. Daniel, L.~M. Maruping, M.~Cataldo, and J.~D. Herbsleb, ``The impact of ideology misfit on open source software communities and companies,'' \emph{MIS Quarterly}, vol.~42, no.~4, p. 1069–1096, 2018.

\bibitem{Gerosa21motivation}
M.~A. Gerosa, I.~Wiese, B.~Trinkenreich, G.~Link, G.~Robles, C.~Treude, I.~Steinmacher, and A.~Sarma, ``The shifting sands of motivation: Revisiting what drives contributors in open source,'' in \emph{Proc. ICSE '21}.\hskip 1em plus 0.5em minus 0.4em\relax {IEEE}, 2021, pp. 1046--1058.

\bibitem{sharp2009models}
H.~Sharp, N.~Baddoo, S.~Beecham, T.~Hall, and H.~Robinson, ``Models of motivation in software engineering,'' \emph{Information and Software Technology}, vol.~51, no.~1, pp. 219--233, 2009.

\bibitem{MockusFH02Coordination}
A.~Mockus, R.~T. Fielding, and J.~D. Herbsleb, ``Two case studies of open source software development: Apache and mozilla,'' \emph{{ACM} Trans. Softw. Eng. Methodol.}, vol.~11, no.~3, pp. 309--346, 2002.

\bibitem{JensenS05}
C.~Jensen and W.~Scacchi, ``Collaboration, leadership, control, and conflict negotiation and the netbeans.org open source software development community,'' in \emph{Proc. HICSS '05}.\hskip 1em plus 0.5em minus 0.4em\relax IEEE, 2005.

\bibitem{StewartG06EffectivenessIdeology}
K.~J. Stewart and S.~Gosain, ``The impact of ideology on effectiveness in open source software development teams,'' \emph{MIS Quarterly}, vol.~30, no.~2, pp. 291--314, 2006.

\bibitem{muller2010grounded}
M.~J. Muller and S.~Kogan, ``Grounded theory method in {HCI} and {CSCW},'' \emph{Cambridge: IBM Center for Social Software}, vol.~28, no.~2, pp. 1--46, 2010.

\bibitem{charmaz2006constructing}
K.~Charmaz, \emph{Constructing Grounded Theory: A Practical Guide through Qualitative Analysis}.\hskip 1em plus 0.5em minus 0.4em\relax SAGE, 2006.

\bibitem{Ljungberg00}
J.~Ljungberg, ``Open source movements as a model for organising,'' \emph{Eur. J. Inf. Syst.}, vol.~9, no.~4, pp. 208--216, 2000.

\bibitem{ZhangWYZLW22Sponsorship}
X.~Zhang, T.~Wang, Y.~Yu, Q.~Zeng, Z.~Li, and H.~Wang, ``Who, what, why and how? towards the monetary incentive in crowd collaboration: A case study of github's sponsor mechanism,'' in \emph{Proc. CHI '22}.\hskip 1em plus 0.5em minus 0.4em\relax ACM, 2022, pp. 206:1--206:18.

\bibitem{GeigerHI21Maintainer}
R.~S. Geiger, D.~Howard, and L.~Irani, ``The labor of maintaining and scaling free and open-source software projects,'' \emph{Proc. {ACM} Hum. Comput. Interact.}, vol.~5, no. {CSCW1}, pp. 1--28, 2021.

\bibitem{QiuDetecting}
H.~S. Qiu, B.~Vasilescu, C.~K\"{a}stner, C.~Egelman, C.~Jaspan, and E.~Murphy-Hill, ``Detecting interpersonal conflict in issues and code review: Cross pollinating open- and closed-source approaches,'' in \emph{Proc. ICSE-SEIS'22}, 2022, p. 41–55.

\bibitem{Dabbish12transparency}
L.~Dabbish, C.~Stuart, J.~Tsay, and J.~Herbsleb, ``Social coding in github: Transparency and collaboration in an open software repository,'' in \emph{Proc. CSCW '12}.\hskip 1em plus 0.5em minus 0.4em\relax ACM, 2012, p. 1277–1286.

\bibitem{Bosu19diversity}
A.~Bosu and K.~Z. Sultana, ``Diversity and inclusion in open source software {(OSS)} projects: Where do we stand?'' in \emph{Proc. ESEM'19}.\hskip 1em plus 0.5em minus 0.4em\relax {IEEE}, 2019, pp. 1--11.

\bibitem{Vasilescu19diversity}
B.~Vasilescu, D.~Posnett, B.~Ray, M.~G. van~den Brand, A.~Serebrenik, P.~Devanbu, and V.~Filkov, ``Gender and tenure diversity in github teams,'' in \emph{Proc. CHI '15}.\hskip 1em plus 0.5em minus 0.4em\relax ACM, 2015, p. 3789–3798.

\bibitem{Canedo20women}
E.~D. Canedo, R.~Bonif{\'{a}}cio, M.~V. Okimoto, A.~Serebrenik, G.~Pinto, and E.~Monteiro, ``Work practices and perceptions from women core developers in {OSS} communities,'' in \emph{Proc. ESEM'20}, M.~T. Baldassarre, F.~Lanubile, M.~Kalinowski, and F.~Sarro, Eds.\hskip 1em plus 0.5em minus 0.4em\relax {ACM}, 2020, pp. 26:1--26:11.

\bibitem{JensenS07Role}
C.~Jensen and W.~Scacchi, ``Role migration and advancement processes in {OSSD} projects: {A} comparative case study,'' in \emph{Proc. ICSE '07}.\hskip 1em plus 0.5em minus 0.4em\relax IEEE, 2007, pp. 364--374.

\bibitem{Trinkenreich20roles}
B.~Trinkenreich, M.~Guizani, I.~Wiese, A.~Sarma, and I.~Steinmacher, ``Hidden figures: Roles and pathways of successful {OSS} contributors,'' \emph{Proc. {ACM} Hum. Comput. Interact.}, vol.~4, no. {CSCW2}, pp. 180:1--180:22, 2020.

\bibitem{HergueuxK22Leadership}
J.~Hergueux and S.~Kessler, ``Follow the leader: Technical and inspirational leadership in open source software,'' in \emph{Proc. CHI '22}.\hskip 1em plus 0.5em minus 0.4em\relax ACM, 2022, pp. 303:1--303:15.

\bibitem{corbin2008basics}
J.~Corbin and A.~Strauss, \emph{Basics of Qualitative Research: Techniques and Procedures for Developing Grounded Theory}.\hskip 1em plus 0.5em minus 0.4em\relax SAGE, 2008.

\bibitem{VanDijk1998}
T.~A. van Dijk, \emph{Ideology: A multidisciplinary Approach}.\hskip 1em plus 0.5em minus 0.4em\relax SAGE, 1998.

\bibitem{Krishnaveni12}
R.~Krishnaveni and R.~Sujatha, ``Communities of practice: An influencing factor for effective knowledge transfer in organizations,'' \emph{IUP Journal of Knowledge Management}, vol.~10, no.~1, pp. 26--40, 01 2012.

\bibitem{SteinmacherGCR19}
I.~Steinmacher, M.~A. Gerosa, T.~U. Conte, and D.~F. Redmiles, ``Overcoming social barriers when contributing to open source software projects,'' \emph{Comput. Support. Cooperative Work.}, vol.~28, no. 1-2, pp. 247--290, 2019.

\end{thebibliography}

\end{document}